# Premature aging as a consequence of Mis-construction of tissues and organs during body development

Jicun Wang-Michelitsch[1]*, Thomas M Michelitsch[2]

[1]Department of Medicine, Addenbrooke's Hospital, University Cambridge, UK (work address until 2007)

[2]Institut Jean le Rond d'Alembert (Paris 6), CNRS UMR 7190 Paris, France

## Abstract

Hutchinson–Gilford Progeria syndrome, Werner syndrome, and Cockayne syndrome are three genetic disorders, in which the children have premature aging features. To understand the phenomenon of premature aging, the similarity of aging features in these syndromes to that in normal aging is investigated. Although these three syndromes have different genetic backgrounds, all the patients have abnormal structures of tissues/organs like that in normal aging. Therefore, the abnormality in tissue structure is the common point in premature aging and normal aging. This abnormality links also a defective development and a defective repair, the Misrepair. Defective development is a result of Mis-construction of the structure of tissues/organs as consequence of genetic mutations. Aging is a result of Mis-**re**constructions, the Misrepairs, for maintaining the structure of tissues/organs. Construction-**re**construction of the structure of an organism is thus the coupling point of development and aging. Mis-construction and Mis-**re**construction (Misrepair) are the essential processes for the development of aging-like feathers. In conclusion, premature aging is a result of Mis-construction of tissues and organs during body development as consequence of genetic disorders.

*email : thomasjicun@gmail.com

Premature aging, called also accelerated aging, is the phenomenon that a child can have aging feathers because of genetic diseases. The three known premature aging syndromes on human being are Hutchinson–Gilford Progeria Syndrome (HGPS), Werner syndrome (WS), and Cockayne syndrome (CS). These syndromes have differences on genetic background, on age of onset of abnormity, and on symptoms. However, they are common on some typical aging changes, including hair loss, tooth loss, thinness and hardness of skin, skin wrinkles, and senior spots. So far, none of traditional aging theory is able to interpret the phenomenon of premature aging. To understand normal aging, we have proposed a generalized concept of Misrepair in Misrepair-accumulation theory (Wang, 2009). The new concept of Misrepair is defined as *Mis-**re**construction of an injured living structure such as a molecule, a cell, and a tissue*. After comparing the premature aging feathers with the changes in normal aging, we find out that premature aging may be a result of Mis-construction of the structure of tissues/organs during body development. In the present paper, we will discuss the relationship between premature aging and normal aging. Our discussion tackles the following issues:

I. Symptoms of premature aging syndromes

II. A novel aging theory: Misrepair-accumulation theory

III. Premature aging as a consequence of Mis-construction of tissues/organs during development

    3.1 Mis-construction and Mis-**re**construction of a tissue/organ
    3.2 Premature aging in animal models by genetic modifications

IV. Analysis of premature aging syndromes

    4.1 Hutchinson-Gilford Progeria syndrome
    4.2 Werner syndrome
    4.3 Cockayne syndrome

V. Conclusions

**I. Symptoms of premature aging syndromes**

HGPS, called also Progeria, is the first disease that is named as premature aging syndrome. HGPS is a genetic condition resulting in abnormal body development and appearance of premature aging feathers since infancy (James, 2005). So far more than 100 cases of HGPS have been reported in the world (Hennekam, 2006). Progeria children often look normal at birth but manifest abnormal growth soon after birth. Growth abnormalities develop progressively with age, including growth failure, hair loss, hardening and thinness of skin, wrinkled skin, stiff joints, atherosclerosis, and loss of body fat and muscle. HGPS children all have a small body but big head with prominent eyes and narrow face. They die mainly from heart attack or stroke in young ages. Differently from that in WS and CS, HGPS children have normal mental development, and they do not develop cancer, cataract, and osteoarthritis. A mutation in *LMNA* gene is thought to be associated with HGPS development (Eriksson, 2003). Normal protein lamin A coded by *LMNA* gene is a part of structure of nuclear lamina and nuclear skeleton in a cell.

WS, called also "adult progeria", is a genetic disorder characterized by an early and progressive development of aging features (James, 2005). WS patients are mainly found in Japan. They have a normal development in childhood, but impaired growth begins after adolescence. The early features of aging are: loss of hair, changing of voice, and hardiness of skin. With time, typical WS symptoms appear and develop, including growth arrest, lens cataracts, skin ulcers, type II diabetes, loss of fertility, hardening of arterial wall, atherosclerosis, osteoporosis, rigidness of joint, and multiple and rare cancers. WS patients often die from cancer or atherosclerosis in their forties or fifties (Ozgenc, 2005). A mutation in *WRN* gene was identified in WS patients. Normal WRN protein is a DNA helicase, and it assists DNA duplication and maintains the functionality of telomeres.

CS, called also "Weber-Cockayne syndrome" and "Neill-Dingwall syndrome", is a rare genetic condition. CS is characterized by growth failure, impaired development of neural system, abnormal sensitivity to sunlight, and appearance of premature aging. Typical aging features in CS include thinness of skin, hair loss, sunken eyes, tooth decay, and hearing loss. CS individuals have small figure with small head (bird head) but long limbs, large hands, and large feet. They often stand in a 'horse-riding stance' because of joint contractures in knees. By the severity of symptoms and the age of onset of symptoms, CS is classified into three subtypes. Type I, called classical type, is the most common subtype, in which the children have abnormal development since age 1. Type II is the severest subtype, and the babies are born with symptoms and die in childhood. Type III is the mildest subtype, in which the children have symptoms only in later childhood. Individuals of type I can survive till twenties, whereas those of type II die before age 7. Pulmonary infection is often the cause of death of CS patients (James, 2005). Gene mutation of a DNA repair enzyme, ERCC6 (8), was identified in CS patients.

## II.  A novel aging theory: Misrepair-accumulation theory

Aging of an organism is characterized by a gradual deformation of structure and reduction of functionality of tissues/organs. We found out that incorrect repair is a common change in different types of aging symptoms, including fibrosis, atherosclerotic plaques, and wrinkles. Like that in Misrepair of DNA and that in scar formation, an incorrect repair will take place when a complete repair of an injury is impossible to achieve. On this basis, we proposed a generalized concept of Misrepair in our Misrepair-accumulation theory (Wang, 2009). The new concept of Misrepair is defined as ***incorrect reconstruction of an injured living structure.*** This concept is applicable to all living structures including molecules (DNAs), cells, and tissues. When an injury is severe, Misrepair, a repair with altered materials and in altered remodeling, is essential for maintaining the structural integrity and for surviving of an organism. Without Misrepairs, an individual could not survive till reproduction age. Misrepair mechanism is thus essential for the survival of a species. However, a Misrepair results in a permanent change of a structure, which can accumulate. Thus, accumulation of Misrepairs disorganize gradually the structure of a molecule, a cell or a tissue, appearing as aging of it. Aging is a side-effect of survival of an organism, but it is beneficial for the survival of the species.

Accumulation of Misrepairs in part of a tissue is not homogeneous but focalized. A consequence of Misrepair is that the misrepaired tissue has increased damage-sensitivity and reduced repair-efficiency. Thus, Misrepairs have a tendency to occur to the part of tissue and its neighborhood where an old Misrepair has taken place. A Misrepair results in the occurrence of more Misrepairs to the same part of a tissue by a viscous circle. Hence, accumulation of Misrepairs is focalized and self-accelerating. The process of aging is therefore self-accelerating. Aging takes place on the levels of molecule, cell, and tissue, respectively; however aging of a multi-cellular organism takes place essentially on tissue level. Aging of an organism and aging of a tissue do not always require aging of cells and aging of DNAs. An irreversible change on the spatial relationship between cells/ECMs in a tissue is *essential and sufficient* for causing a decline of organ functionality. In summary, aging of an organism is a result of accumulation of Misrepairs on tissue level.

## III. Premature aging as a consequence of Mis-construction of tissues/organs during development

Why do the children with these syndromes have the symptoms that are similar to that in normal aging? To answer this question, one needs to search for the common change underlying both of normal aging and premature aging. In our view, these children have aging features because they have abnormal structures of tissues/organs. Thus, the abnormality in tissue structure is the common point in premature aging and normal aging. Interestingly, this abnormality links also a defective development and a defective repair, the Misrepair.

### 3.1 Mis-construction and Mis-reconstruction of a tissue/organ

Development and repair are both a process of constructing of the structure of a tissue/organ. Any element that interrupts the constructing process, including the processes of cell division, cell organization, ECMs-production, and ECMs-modeling, will affect the result of development and repair. Abnormal and defective development is thus a result of *Mis-construction* of the tissues/organs of an organism. Defective development will lead to low functionality of the tissues/organs. Aging is a result of accumulation of *Mis-reconstructions* (Misrepairs) of tissues/organs. Thus, construction-**re**construction of the structure of an organism is the coupling point between development and aging. In this aspect, aging and development are regulated by the same mechanism, as that proposed in the developmental aging theory (Zwaan, 2003). Construction of an organism is genetically controlled, whereas reconstruction (repair) of an organism is promoted by an injury. Figure 1 shows the relationship between premature aging and normal aging. In normal development of a tissue, correct construction makes a perfect organization of cells/ECMs, which has full functionality. In premature aging, Mis-construction results in an altered organization of cells/ECMs in the whole tissue, which has low functionality. In normal aging, Misrepair is promoted by death of cells/ECMs, and Mis-**re**construction results in an altered reorganization of cells/ECMs in part of a tissue, which has reduced functionality.

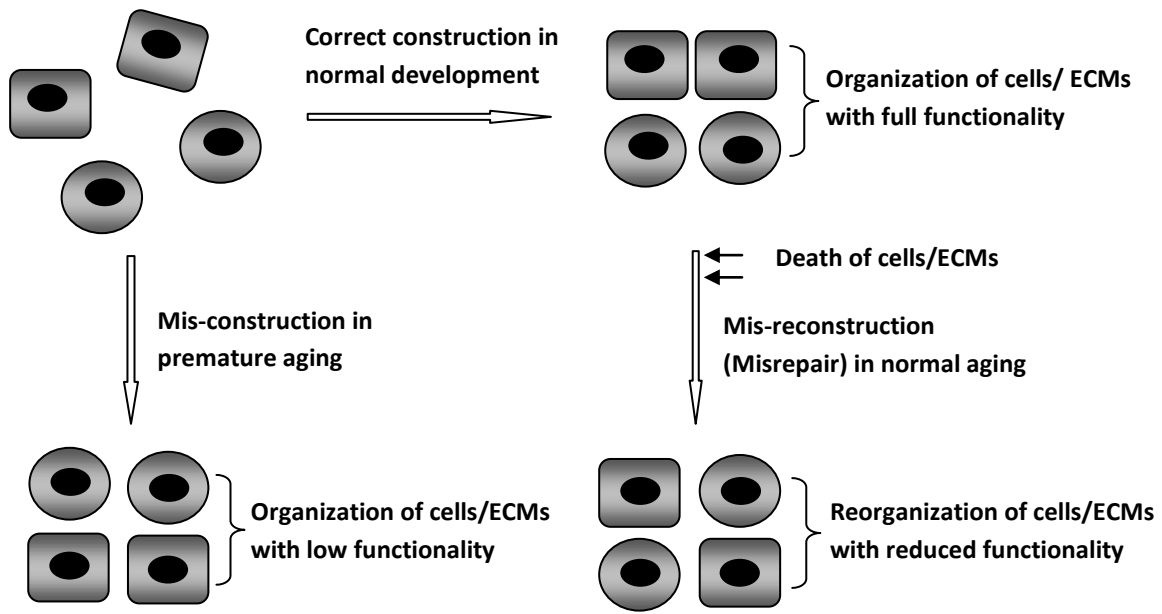

**Figure 1. Abnormality in tissue structure: the common point in normal aging and premature aging**

In normal development of a tissue, correct construction makes a perfect organization of cells/ECMs, which has full functionality. In premature aging, Mis-construction makes an altered organization of cell/ECMs in the whole tissue, which has low functionality. In normal aging, Misrepair is promoted by death of cells/ECMs, and Mis-reconstruction results in an altered reorganization of cells/ECMs in part of a tissue, which has reduced functionality.

Misrepairs are local, only affecting the injured part of a tissue. Differently, the Mis-construction caused by genetic disorders is systemic, affecting all tissues/organs. The organs that develop defectively in these syndromes have low potential of functionality and will go to failure earlier and faster. A Misrepair is not a "real mistake" but a compromise of an organism that is essential for surviving. Differently, the Mis-construction during body development is a "real mistake", and it will lead to failure of the development. For the individuals with premature aging syndromes, especially for those, who have survived over childhood, Misrepairs also contribute to their aging feathers. The phenomenon of premature aging is a powerful evidence to disprove some traditional aging theories, because premature aging features in these syndromes cannot be: **A**. caused by damage (Damage-accumulation theory (Kirkwood, 2005)); **B**. caused by free radicals (Free-radical theory (Harman, 1956)); **C**. controlled by a common gene (Gene-controlling theory (Bell, 2012)); or **D**. due to cell senescence (Cell senescence/telomere theory (Hayflick, 1965; Blackburn, 2000)).

### 3.2 Premature aging in animal models by genetic modifications

Premature aging features are also observed in animal models, in which the animals are genetically modified and have abnormal developments. Protein fibulin-5 is an extracellular molecule and important in assembling and modeling elastic fibers in extracellular matrixes. A mutation in gene *FBLN 5* in human being results in production of non-functional fibulin-5 proteins. This mutation is found to be the cause of cutis laxa, a disease characterized by weakened connective tissues and wrinkled skin. A study with an animal model showed that the mice that have Fibulin-5-/- modification undergo defective development of connective tissues. These mice manifest typical aging symptoms, such as loose and wrinkled skin, vascular abnormalities, and emphysema (Hirai, 2007). Protein Klotho is a type of trans-membrane protein with activity of β-glucuronidase. Protein Klotho is important in regulating cell calcium homeostasis. Studies showed that Klotho-deficient mice have multiple accelerated aging changes including extensive and accelerated arteriosclerosis (Kuro-o, 1997). Impaired development of blood vessels is found to be associated with the dysfunction of vasodilatation and angiogenesis in these mice (Lanske, 2007). All these studies on different animal models and with different genetic modifications reveal one thing in common: a defective development by genetic changes can result in premature aging features (Sun, 2004; Razague, 2006; Trifunovic, 2004; de Boer, 2002; Espada, 2008).

## IV.  Analysis of premature aging syndromes

Although the gene mutations have been identified in these syndromes, concrete associations of these mutations with development of syndromes are not clear. In this part, we analyze the effects of these gene mutations on body development and try to explain some specific symptoms in each syndrome.

### 4.1  Hutchinson-Gilford Progeria syndrome

A characteristic pathological change in HGPS is the blebbing of cell nucleus (Lans, 2006). The abnormal structure of nuclear lamina caused by mutation of *LMNA* gene is thus thought to be the origin for HGPS development (Eriksson, 2003). Nuclear lamina is part of the structure of nuclear skeleton, and it is important in maintaining nuclear shape and maintaining the functional organization of chromosomes in nucleus. Nuclear lamina is composed of lamin A and lamin B. For constructing nuclear lamina, protein lamin A, in form of pre-lamin A, is transported from cytoplasm and localized to the inner side of nuclear membrane by its farnesyl group. After localization, the farnesyl group will be cut off, and the free lamin A can be integrated into nuclear lamina. In HGPS, mutation of *LMNA* gene results in production of a shorter form of pre-lamin A. Lack of the 50 amino acids at carboxyl-terminus, the mutant lamin A, called progerin, loses its cleaving site for cutting-off the farnesyl group. As a consequence, the mutant lamin A cannot be free from nuclear membrane and integrated into nuclear lamina. Lacking of lamin A, the structure of nuclear lamina is defective and fragile, and the shape of nucleus is altered. In addition, the non-functional proteins of lamin A accumulate beneath nuclear membrane. Defective structure of nuclear lamina and deposition of lamin A proteins may both contribute to the blebbing change of nucleus in HGPS (Dechat, 2007). Nuclear functions, including organization of chromosomes, DNA duplication, cell mitosis, RNA transcription, and substance transportation, can be all affected in the cells of HGPS by the defect of nuclear lamina.

However, the concrete link between the defect of nuclear lamina and the symptoms of HGPS is not known. Apart from HGPS, some other diseases are also related to gene mutations of lamin A or lamin C, and they include Emery-Dreifuss muscular dystrophy (Madej-Pilarczyk, 2016) and dilated cardiomyopathy (Wang, 2017). Deformation of nuclear membrane is also seen in these dystrophies; however most of the patients with these diseases can survive over middle age. A common pathology in these diseases is the fragility of the muscular cells in skeleton and in the heart. The patients have often muscular dystrophy from death of muscular cells during cell deformation. A defect of nuclear lamina caused by mutations of lamin A/C might account for the fragility of muscular cells in these diseases. In our view, HGPS children may have also fragility on muscular cells and other cells due to defect of nuclear lamina. The fragility of muscular cells may be a reason why HGPS children die often from heart failure. The defect of nuclear lamin in HGPS might be severer than that in the above diseases. In HGPS, the cell death caused by cell fragility and cell deformations may occur to not only muscular cells but also other types of cells that need to deform during functioning, such as fibroblasts and endothelium cells. Cell death of these types of cells during body development may severely affect the construction of tissue/organ structures, and this may account for the defective development in HGPS. Exceptionally, the neuron cells in the brain need not deform for functioning. This may explain why the development of neural system is not severely affected in HGPS.

## 4.2  Werner's syndrome

*WRN* mutation is thought to be associated with WS development. WRN protein is a kind of DNA helicase and it is important in protecting telomeres. *WRN* gene is inherited in autosomal recessive pattern, thus only those individuals who have two copies of mutated *WRN* gene can develop WS. The WRN mutant in WS is a shortened variant of WRN protein, and it loses the ability of interacting with DNA and protecting telomeres (Gray, 1997). Without the protection by WRN protein, telomeres may be damaged in nucleus and lose their functionality on stabilizing chromosomes. The subsequent chromosome instability will lead to dysfunctions of nucleus and cells, including failure of DNA duplication, adoption of DNA mutations, failure of separation of chromosomes during cell division, and failure of gene expression. Destination of a cell with these dysfunctions can be one of the followings: **A**. cell death as a result of gene mutations or failure of gene expression, **B**. failure of cell division as a result of failure of DNA duplication or failure of separation of chromosomes, and **C**. cell transformation as a result of accumulation of adopted DNA mutations. Cell death, failure of cell division, and cell transformation may all contribute to the defective body development in WS.

Interestingly, some symptoms that develop in WS do not develop in HGPS and in CS. Specific symptoms in WS include retardation of onset of symptoms, occurrence of multiple and rare cancers, and development of lens cataract. In our view, the retardation of onset of symptoms in WS might be due to a delayed effect of WNR mutation. In the cells that have mutant WRN protein, the telomeres may be shortened more rapidly than normal.  The defective body development in WS may only start when the telomeres in some cells are too short and lose their functionality.  Cancer-development in WS might be a result of accumulation of adopted DNA mutations. Genomic instability makes WS cells adopt new

DNA mutations in high frequency. Important are: **A.** new DNA mutations can occur to any one of the cells of WS patients; and **B.** the defective cells may have high tolerance to DNA mutations. Rapid accumulation of DNA mutations in many cells may account for the occurrence of multiple and rare tumors in WS. Development of lens cataract in WS might be a result of failure of cell division of epithelial cells in the lens capsule. Transparency of the lens relies on the regular arrangement of lens fibers. Lens fibers are derived from the epithelial cells in lens capsule. These epithelial cells need to divide constantly for producing new lens fibers for maintaining the regular arrangement of lens fibers. However, in WS, failure of cell division can lead to failure of maintenance of the transparency of the lens.

### 4.3  Cockayne's syndrome

Gene mutations of a DNA repair enzyme were found to be associated with CS development. This enzyme is coded by *ERCC6* gene and *ERCC8* gene. These genes are inherited in autosomal recessive pattern, and the CS patients all have two copies of mutated genes (Bertola, 2006). In CS, the mutation of *ERCC6(8)* gene results in a defective form of ERCC6 (8) protein, which loses its ability of repairing certain types of DNA injuries (Hoeijmakers, 2009). The DNAs in a cell can be injured in different manners, and different groups of enzymes are responsible for repairing different types of DNA injuries. ERCC6 (8) protein is an enzyme essential for the transcription-coupled nucleotide excision repair (TC-NER). Thus, a defect of DNA repair caused by *ERCC6 (8)* mutation can make all somatic cells in danger of death from DNA injuries. In our view, cell death from DNA injuries in early stage of body development as consequence of defect of DNA repair is responsible for CS development.

**Firstly,** death of cells in early stage of body development will affect severely the development. In CS, due to failure of DNA repair, the cells that suffer from DNA injuries may die during embryonic development. Most of embryonic cells have the potential of proliferation and differentiation; and death of some cells can severely impair the further development. It is known that different patients with CS have great difference on the severity of symptoms. In our view, the severity of CS is more likely related to the age of onset of cell death. The earlier the cell death takes place, the severer it will affect the development. **Secondly,** CS children have severe impaired development of nerve system. A normal baby can have the same number of neurons at birth as an adult. A child of three years old has the same weight of the brain as an adult. This means that in mental development production of neuron cells takes place mainly before birth. Thus, cell death in early stage of body development can severely impair mental development. **Thirdly,** CS children have abnormal sensitivity to sunlight. In nature, DNA injuries are mainly caused by UV radiation in sunlight. The sensitivity of CS cells to DNA injuries because of defect of DNA repair is in our view the main reason for the hyper-sensitivity of CS patients to sunlight.

Another interesting phenomenon in CS is that the patients do not develop cancer. In our view, cell transformation is a result of accumulation of DNA mutations through many generations of cells (Wang-Michelitsch, 2015). Two preconditions for accumulation of DNA mutations are: **A**. survival of the cells that have DNA mutations, and **B**. potential of cell proliferation. A DNA mutation in a somatic cell occurs as a result of Misrepair of injured DNA. However,

Misrepair of DNA can only take place when the cells have normal functionality on DNA repair. In CS, the cells have defect on DNA repair; therefore they have no ability to make complete repair and Misrepair of DNA! The chance of accumulation of DNA mutations and cell transformation in CS is too low!

**V.   Conclusions**

Premature aging syndromes are a group of genetic diseases in which the children develop aging feathers. The premature aging seen in these disorders is not a result of aging rather a result of defective development. The abnormality in tissue structure is the common point in normal aging and premature aging. The defective development in these syndromes results from Mis-construction, whereas normal aging results from Misrepairs (Mis-**re**constructions). Construction-reconstruction of the structure of an organism is thus the coupling point of development and aging. Mis-construction and Mis-**re**construction (Misrepair) are the essential processes for the development of aging-like feathers.